\documentclass[12pt,preprint]{aastex}

\begin{document}

\title{General Relativistic Magnetohydrodynamic Simulations of Collapsars: Rotating Black Hole Cases}
    
\author{Yosuke Mizuno\altaffilmark{1}, Shoichi Yamada\altaffilmark{2}, Shinji Koide\altaffilmark{3} and Kazunari Shibata\altaffilmark{4}}

\altaffiltext{1}{Department of Astronomy, Kyoto University, Sakyo, 
    Kyoto 606-8502, Japan; mizuno@kusastro.kyoto-u.ac.jp}
\altaffiltext{2}{Science and Engineering, Waseda University, Shinjuku, 
    Tokyo 169-8555, Japan; shoichi@heap.phys.waseda.ac.jp}
\altaffiltext{3}{Department of Engineering, Toyama University, Gofuku, 
Toyama 930-8555, Japan; koidesin@ecs.toyama-u.ac.jp}
\altaffiltext{4}{Kwasan and Hida Observatory, Kyoto University, Yamashina, 
    Kyoto 607-8471, Japan; shibata@kwasan.kyoto-u.ac.jp}
    
\shorttitle{GENERAL RELATIVISTIC MHD SIMULATIONS OF COLLAPSARS}

\begin{abstract}

We have performed 2.5-dimensional general relativistic magnetohydrodynamic (MHD) simulations of collapsars including a rotating black hole. This paper is an extension of our previous paper (Mizuno et al. 2004). The current calculation focuses on the effect of black hole rotation using general relativistic MHD with simplified microphysics, i.e., we ignore neutrino cooling, physical equation of state and photodisintegration. Initially, we assume that the core collapse is failed in this star. A few $M_{\odot}$ rotating black hole is inserted by hand into the calculation. We consider two cases, the co-rotating case and counter-rotating case with respect to the black hole rotation. Although the counter-rotating case may be unrealistic for collapsar, we perform as a maximally dragging case of magnetic field. The simulation results show the formation of a disk-like structure and the generation of a jet-like outflow near the central black hole. The jet-like outflow propagates outwardly with the twisted magnetic field and becomes collimated. We have found that the jets are generated and accelerated mainly by the magnetic field. The total jet velocity in the rotating black hole case is comparable to that of the non-rotating black hole case (Mizuno et al. 2004), $\sim$ 0.3c. When the rotation of the black hole is faster, the magnetic field is twisted strongly owing to the frame-dragging effect. The magnetic energy stored by the twisting magnetic field is converted to kinetic energy of the jet directly rather than propagating as an Alfv\'{e}n wave. Thus, as the rotation of the black hole becomes faster, the poloidal velocity of the jet becomes faster. In the rapidly rotating black hole case the jet-like outflow can be produced by the frame dragging effect only through the twisting of magnetic field even if there is no stellar rotation.

\end{abstract}

\section{Introduction}

It is now generally believed that ``long-soft'' gamma-ray bursts (GRBs) are a phenomenon related to the deaths of massive stars. There are direct and indirect observational evidence of the close relationship between GRBs and supernovae, such as the observed association with star-forming regions in galaxies (Vreeswijk et al. 2001; Bloom, Kulkarni, \& Djorgovski 2002; Gorosabel et al. 2003), the ``bump'' observed in the afterglows of some GRBs (Reichert 1999; Galama et al. 2000; Bloom et al. 2002; Garnavich et al. 2003; Bloom et al. 2003), metal emission lines observed in the X-ray afterglow of GRB011211 (Reeves et al. 2002), and  the association of GRB980425 with SN 1998bw (Galama et al. 1998b). Recently, another clear evidence was found, GRB030329 was accompanied by a bright energetic supernova of Type Ic, SN 2003dh (Price et al. 2003b; Hjorth et al. 2003; Stanek et al. 2003). It has been suggested so far that some of GRBs are produced when the iron core of a massive star collapses either to a black hole (Woolsey 1993; MacFadyen \& Woosley 1999) or to a rapidly rotating highly magnetic neutron star (Wheeler et al. 2000), both of which eventually produce a relativistic jet.

The collapsar model is one of the most promising scenarios involving massive stars (Woosley 1993; MacFadyen \& Woosley 1999). Collapsar is a rotating massive star, lacking a hydrogen envelope. In this model, the iron core of the rotating massive star collapses to a black hole surrounded by an accretion disk. The accretion through this disk produces outflows via neutrino annihilation and/or magnetohydrorynamic (MHD) processes. They are further collimated by the passage through the stellar mantle.  The formation and propagation of relativistic flows from collapsar has been studied numerically by both Newtonian (MacFadyen \& Woolsey 1999; MacFadyen, Woosley, \& Heger 2001) and special relativistic hydrodynamic simulations (Aloy et al. 2000; Zhang, Woosley, \& MacFadyen 2003). However, these previous numerical simulations of the collapsar model did not fully address the formation mechanism of the relativistic outflow. 

It is suspected that magnetic fields may play an important role in the formation and acceleration of relativistic jets. In numerical simulations of gravitational collapse of massive stars the effect of stellar rotation and intrinsic magnetic fields was studied by several authors (LeBlanc \& Wilson 1970; Symbalisty 1984; Ardeljan et al. 2000; Kotake et al. 2004; Yamada et al. 2004). Symbalisty (1984) showed the formation of high-density, supersonic jets in the combination of a rapid rotation and a strong dipole magnetic field, which was confirmed by Yamada et al. (2004). 

In our previous paper (Mizuno et al. 2004) we have performed 2.5-dimensional general relativistic MHD simulations of the gravitational collapse of a rotating magnetized massive star with a non-rotating black hole at the center as a model for collapsar. We showed the formation of a disk-like structure and the generation of a mildly relativistic jet ($\sim 0.3 c$) inside the shock wave launched at the core bounce. We have found the jet is accelerated by the magnetic pressure and the centrifugal force and is collimated by the pinching force of the toroidal magnetic field amplified by the rotation and the effect of geometry of the poloidal magnetic field. The Poynting flux transports more energy outward than the jet. However, the jet in these simulations is too slow to applicable for GRBs. In our previous paper, we put a non-rotating black hole at the center, which is not appropriate for the rotational collapse. It is natural that the iron core collapses to a rotating black hole. The rotating black hole (Kerr black hole) is useful to form the relativistic jet for some reasons. If the black hole is rotating, two kinds of energy are available for generating a jet. One is the rotational energy of the accreted matter and the other is the rotational energy of the rotating black hole itself. Blandford and Znajek (1977) investigated magnetospheres of Kerr black holes and derived a force-free solution for the electromagnetic field. Their results showed that electromagnetic energy is radiated from the black hole horizon directly. This is called Blandford-Znajek mechanism. However, the direct energy emission from the black hole horizon appears to be inconsistent with the causality at the horizon.  Punsly and Coroniti (1990) found that if the ergospheric plasma gets frozen onto large-scale magnetic field lines, it can drive a magnetic wind to infinity. Using a general relativistic MHD code, the basic mechanism of energy extraction from the Kerr black hole via magnetic field is investigated numerically (Koide et al. 2002; Koide 2003). The numerical results showed that the rotational energy of the Kerr black hole can be extracted when the magnetic field is strong enough.

Here, as a collapsar model we perform 2.5-dimensional general relativistic MHD simulations of the gravitational collapse of a rotating magnetized massive star with a rotating black hole at the center. We investigate the physics of the formation of jets, the acceleration force on the jets, and the dependence on the rotation parameter of the black hole.  This paper is extension of our previous paper (Mizuno et al. 2004). We describe the numerical method of our simulations briefly in section 2 and present our results in section 3. The summary and discussion are given in section 4.

\section{Numerical Method}

\subsection{Basic Equations}

In order to study the formation of relativistic jets from collapsar we use a 2.5-dimensional general relativistic magnetohydrodynamics (GRMHD) code with Boyer-Lindquist coordinates $(R, \theta, \phi)$ (Koide 2003, Mizuno et al. 2004). The method is based on a 3+1 formalism of the general relativistic conservation laws of particle-number and energy-momentum, Maxwell equations, and Ohm's law with no electric resistance (ideal MHD condition) on a curved space-time (Thorne, Price, \& Macdonald 1986; Koide et al. 2000; Koide 2003; Mizuno et al. 2004).

The space-time $(x^{0}, x^{1}, x^{2}, x^{3})=(ct,x^{1},x^{2},x^{3})$ is described by the metric $g_{\mu \nu}$, where the line element $ds$ is given by $(ds)^{2}= g_{\mu \nu} dx^{\mu}dx^{\nu}$. Here, $c$ is the speed of light.
If we assume that the off-diagonal spatial elements of the metric $g_{\mu \nu}$ vanish, 
\begin{equation}
g_{i j} = 0 \quad (i \neq j),
\end{equation}
and we use the notation,
\begin{equation}
g_{00} = -h_{0}^{2}, \quad g_{ii}=h_{i}^{2},
\end{equation}
\begin{equation}
g_{i0} = g_{0i} = -h_{i}^{2}\omega_{i}/c, 
\end{equation}
then the line element can be written
\begin{eqnarray}
(ds)^{2} &=& g_{\mu \nu} dx^{\mu} dx^{\nu} \\
&=& -h^{2}_{0} (cdt)^{2} + \sum^{3}_{i=1} [h^{2}_{i}(dx^{i})^{2}-2 h^{2}_{i}\omega_{i}dtdx^{i}].
\end{eqnarray}
When we define the lapse function $\alpha$ and ``shift velocity'' (shift vector) $\beta^{i}$ as 
\begin{equation}
\alpha = \sqrt{h_{0}^{2} + \sum^{3}_{i=1} \left( {h_{i} \omega_{i} \over c}\right)^{2}},
\end{equation}
\begin{equation}
\beta^{i} = {h_{i} \omega_{i} \over c \alpha},
\end{equation}
the line element $ds$ is written as
\begin{equation}
(ds)^{2} = - \alpha^{2} (cdt)^{2} + \sum^{3}_{i=1} (h_{i} dx^{i} - c \beta \alpha dt)^{2}.
\end{equation}
The contravariant metric is written explicitly as
\begin{equation}
g^{00} = - {1 \over \alpha^{2}},
\end{equation}
\begin{equation}
g^{i0} = g^{0i} = -{1 \over \alpha^{2}}{\omega_{i} \over c},
\end{equation}
\begin{equation}
g^{ij} = {1 \over h_{i} h_{j}} (\delta^{ij}-\beta^{i}\beta^{j}),
\end{equation}
where $\delta^{ij}$ is the Kronecker's $\delta$ symbol.

In the GRMHD code, a simplified total variation diminishing (TVD) method is employed (Davis 1984). This method is similar to the Lax-Wendroff method with addition of a diffusion term and is useful because it requires only the maximum speed of physical waves but not each eigenvector or eigenvalue of the coefficient matrix of the linearized GRMHD equations.

We do not consider the evolution of the metric because the accreted mass is sufficiently small on the time scale of the simulations. 
Previous works of collapsar simulation included microphysics. MacFadyen \& Woosley (1999), for example, took into account a physical equation of state (EOS), photodisintegrations and neutrino cooling. Proga et al. (2003) implemented them in MHD. In this paper we neglect these microphysics entirely and concentrate on the general relativistic magnetohydrodynamics, particularly the effect of the rotation of black hole. We assume mainly for numerical simplicity that matter can be described as an ideal gas with a gamma law EOS ($p \propto \rho^{\Gamma}$) although we know that the gamma law EOS is not a very good approximation for the gravitational collapse of massive stars. We further assume that $\Gamma = 5/3$ in the simulations for numerical reasons although it is more appropriate to adopt $\Gamma = 4/3$ for the gas of current interest.

\subsection{Initial Condition}

As for the initial model, we have the collapsar model in mind. In principle, we should start calculations from a realistic progenitor model with rotation and magnetic field. We take, however, the following pragmatic approach as a rough guide, we use the post bounce profile of Bruenn's realistic 1D supernova model (Bruenn 1992). As a supernova model, this model is failed. Nominally believed, this model will produce a black hole later. Since our computation is scale free, we employ only the profiles of the density, pressure and radial velocity as our initial condition. Furthermore we put a few $M_{\odot}$ black hole at the center. The initial condition in this paper is the same as our previous paper except for the metric providing the background space-time (see Mizuno et al. 2004). In this way, we can discuss generic features of the dynamics. 

A rotating black hole (Kerr black hole) has two characteristic parameters: Its mass $M$ and angular momentum $J$. We often use the rotation parameter $a=J/J_{\mathrm{max}}$, where $J_{\mathrm{max}}=GM^{2}/c$ is the angular momentum of a maximally rotating black hole with mass $M$. In the Boyer-Lindquist coordinates, the metric of Kerr space-time is written as
\begin{equation}
h_{0} = \sqrt{1-{2 r_{\mathrm{g}}R \over \Sigma}}, \quad h_{1}= \sqrt{{\Sigma \over \Delta}}, \quad h_{2}=\sqrt{\Sigma}, \quad h_{3}=\sqrt{{A \over \Sigma}} \sin\theta,
\end{equation}
\begin{equation}
\omega_{1} = \omega_{2} = 0, \quad \omega_{3}= {2 c r^{2}_{g} a R \over A}
\end{equation}
where $r_{\mathrm{g}} \equiv GM/c^{2}$ is the gravitational radius, $\Delta = R^{2} - 2 r_{\mathrm{g}} R + (a r_{\mathrm{g}})^{2} \cos^{2} \theta$, $\Sigma = R^{2} + (a r_{\mathrm{g}})^{2} \cos^{2} \theta$ and $A = \{ R^{2} + (ar_{\mathrm{g}})^{2} \}^{2}- \Delta (a r_{\mathrm{g}})^{2} \sin^{2} \theta$. In this metric, the lapse function is $\alpha = \sqrt{\Delta \Sigma /A}$. The radius of the event horizon is $r_{\mathrm{H}} = r_{\mathrm{g}} (1+ \sqrt{1-a^{2}})$, which is found by setting $\alpha=0$. We also use the Schwarzschild radius of the black hole, $r_{\mathrm{S}} = 2 GM / c^{2} = 2 r_{\mathrm{g}}$ as a unit of length in this paper.

We add by hand the stellar rotation and intrinsic magnetic field to the originally non-rotating , non-magnetic model as an initial condition. The initial distribution of rotational velocity is assumed to be a function of the distance from the rotation axis, $r = R \sin \theta$, only:
\begin{equation}
v_{\phi} = v_{0} {x_{0}^{2} \over r^{2} + x_{0}^{2}} r.
\end{equation}
Here, $v_{0}$ is a model parameter for rotational velocity. We fix $x_{0} = 100 \ r_{\mathrm{S}}$ in this paper. We include the stellar rotation only up to 18 $r_{\mathrm{S}}$. Because the accreted mass for stellar matter in pre-shock region is sufficiently small on the time scale of our interest, the neglect of rotation in the pre-shock region does not affect the dynamics. The rotation profile is similar to the previous simulations for the gravitational collapse of a rotating core (M\"{o}nchmeyer \& M\"{uller 1989; Yamda \& Sato 1994}).
The initial magnetic field is assumed to be uniform and parallel to the rotational axis. This is known as the Wald solution (Koide 2003). We use $B_{0}$ as a model parameter for magentic field strength.

We emphasize that our simulations are scale free. The normalization units and typical values for normalizations are found in Mizuno et al. (2004).
The models computed in this paper are summarized in Table {\ref{table1}}.

We use the Zero Angular Momentum Observer (ZAMO) system for the 3-vector quantities, such as velocity $\mathbf{v}$, magnetic field $\mathbf{B}$, electric current density $\mathbf{J}$, and so on. For scalars, we use the frame comoving with the fluid flow.
The simulations are done in the region 1.4 $r_{\mathrm{S}}$ (KA1, KA2, KB7, KC8), 1.6 $r_{\mathrm{S}}$ (KB6), 1.8 $r_{\mathrm{S}}$ (KB4, KB5), 2.0 $r_{\mathrm{S}}$ (KB3) $ < R < $ 60 $r_{\mathrm{S}}$, $0 < \theta < \pi/2$ with $120 \times 120$ mesh points. We assume axisymmetry with respect to the z-axis and mirror symmetry with respect to the equatorial plane. We employ a free boundary condition at the inner and outer boundaries of the radial direction.

\section{Numerical Results}

\subsection{Formation of the Jet}

\begin{figure}
\plotone{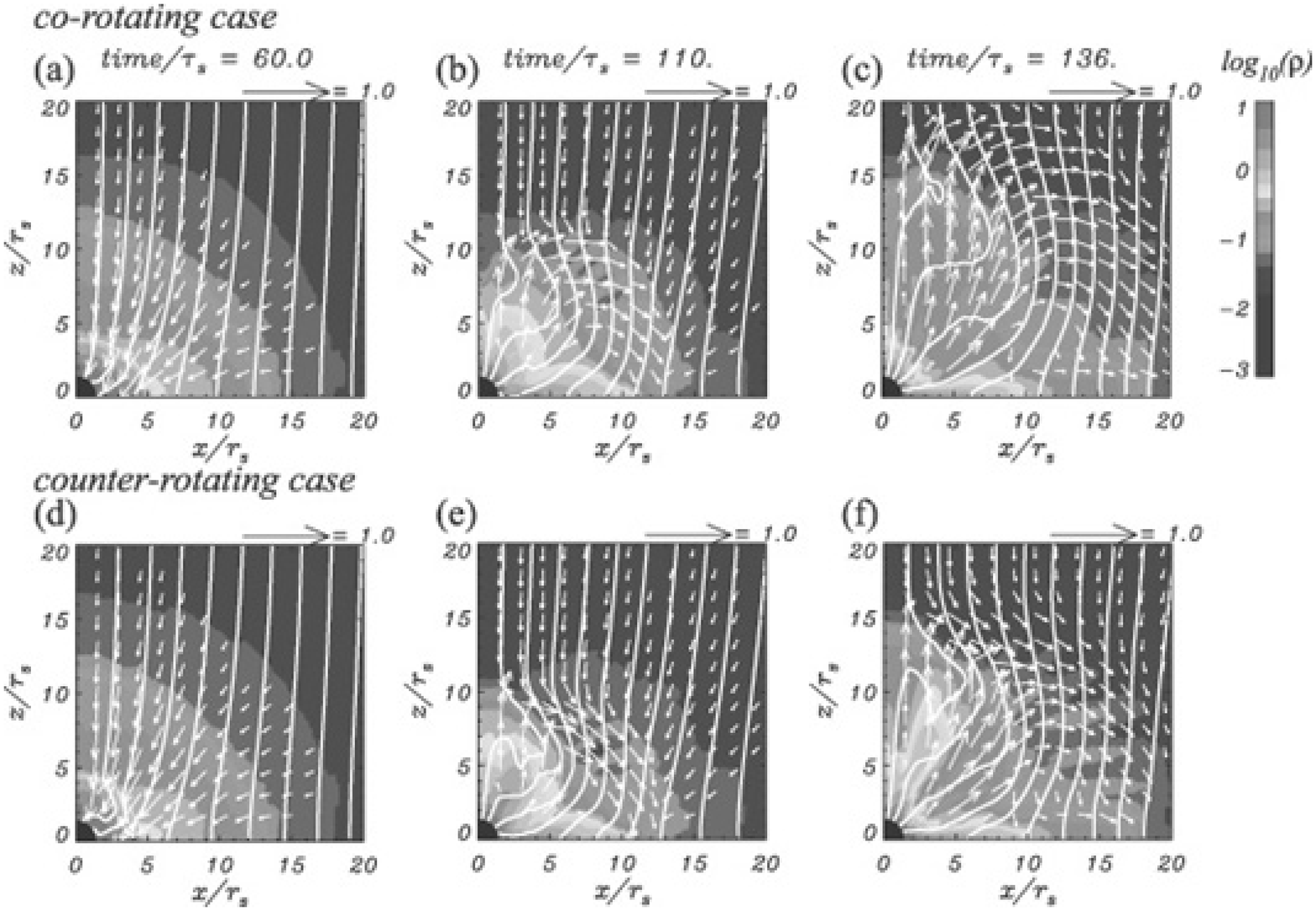}
\caption{The time evolution of the density for the co-rotating case KA1 (upper panels) and the counter-rotating case KA2 (lower panels). The color scale shows the value of the logarithm of density. The white curves depict magnetic field lines. Arrows represent the poloidal velocities normalized by the light velocity. The central black region corresponds to a black hole, though the actual boundary is not located at the surface of the black hole, but at $r/r_{\mathrm{S}}=1.4$ for the numerical reasons. The distance and the time are given in units of $r_{s}$ and $\tau_{\mathrm{S}} \equiv r_{\mathrm{S}}/c$, respectively. (a), (d) Condition at $t/\tau_{\mathrm{S}} =60$. The co-rotating and the counter-rotating cases show  almost the same evolution. The stellar matter falls and rotates around the black hole. Accreting stellar matter forms a disk-like structure in the equatorial plane. (b), (e) Condition at $t/\tau_{\mathrm{S}} = 110$. In both cases, the jet-like outflow is produced near the black hole. (c), (f) Condition at $t/\tau_{\mathrm{S}} = 136$. \label{fig1}}
\end{figure}

We shall first discuss the results of the co-rotating case KA1 (Table \ref{table1}; from now on referred to as the standard case) and the counter-rotating case KA2 with respect to the black hole rotation. Although the counter-rotating case may be unrealistic for collapsar, we performed as the maximally dragging case of magnetic field. The model parameters are $a = 0.999$ and $-0.999$, $B_{0} = 0.05$, $v_{0}=0.01$.
They are the same as those in the case A2 of Mizuno et al. (2004) except for the rotation parameter. Figure \ref{fig1} shows the time evolution of density for KA1 together with the counter-rotating case KA2. Almost the same evolution is shown in both cases. The stellar matter falls onto the central black hole at first. The matter piles up on the equatorial plane, and forms a disk-like structure. Since the magnetic field is frozen into the plasma, it is dragged and deformed by the accreting matter. At about $t/\tau_{\mathrm{S}}=60$, a shock wave and the jet-like outflow are produced near the central black hole and propagate outward. The jet-like outflow in the case of rotating black hole is {\it more} powerful and is ejected from the more {\it inner} region than that in the case of non-rotating black hole (Mizuno et al. 2004). Although the jet-like outflow in the latter case is ejected from the region away from the axis of rotation by about $r/r_{\mathrm{S}} = 5$ and the matter near the rotation axis fall to the central black hole, the jet-like outflow in the former case is ejected closer to the rotation axis. Comparing the jet-like outflow for the rotating black hole with the one in the previous simulations, 
we find that our jets are similar to these in the Newtonian MHD simulations (LeBlanc \& Wilson 1970; Symbalisty 1984), which shows the formation of jet-like outflow close to the rotation axis. The jet in the pseudo-Newtonian MHD simulations (Proga et al. 2003) is more similar to the jet for the non-rotating black hole case. We emphasize that the generation mechanism of jets in those MHD simulations are common, MHD effects.

\begin{figure}
\plotone{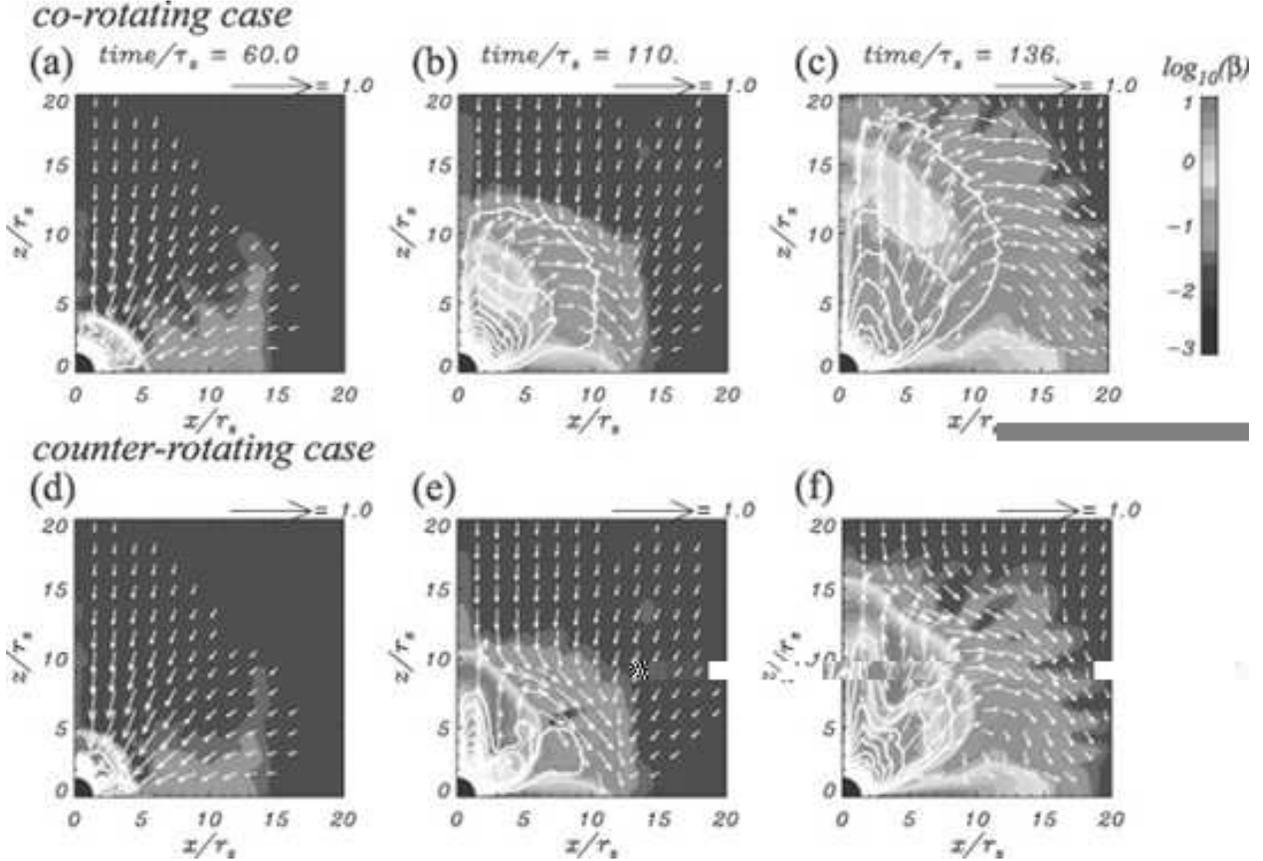}
\caption{The time evolution of the plasma beta ($\beta = P_{\mathrm{gas}}/P_{\mathrm{mag}}$) distribution for the co-rotating case KA1 (upper panels) and the counter-rotating case KA2 (lower panels). The color scale shows the value of the logarithm of plasma beta. The white contour plots are the toroidal component of the magnetic field. The contour level step-width is 0.05 for (a), (b), and (c), 0.067 for (d), (e), and (f) in units of the toroidal magnetic field ($B_{\phi}$). Arrows represent the poloidal velocities normalized by the light velocity. The central black region corresponds to a black hole, though the actual boundary is not located at the surface of the black hole, but at $r/r_{\mathrm{S}}=1.4$ for numerical reasons. The distance and the time are given in units of $r_{\mathrm{S}}$ and $\tau_{\mathrm{S}} \equiv r_{\mathrm{S}}/c$, respectively. The amplified magnetic field expands with a jet-like outflow. \label{fig2}}
\end{figure}

Figure \ref{fig2} shows the time evolutions of plasma beta ($\beta = P_{\mathrm{gas}}/P_{\mathrm{mag}}$, where $P_{\mathrm{gas}}$ is gas pressure and $P_{\mathrm{mag}}$ is magnetic pressure) and toroidal magnetic field for the co-rotating case KA1, again with the counter-rotating case KA2. In both cases, the magnetic field is twisted due to the differential rotation of accreted matter and the frame dragging effect of the rotating black hole and is amplified significantly near the central black hole. The amplified magnetic field expands outwards with the jet-like outflow and collimates it. In the case of rotating black hole, it is not clear that the amplified magnetic field propagates as an Alfv\'{e}n wave in contrast to the case of non-rotating black hole but the amplified magnetic field launches an outgoing shock wave. 
 The plasma beta distribution is a little different between the co-rotating and the counter-rotating cases. In the co-rotating case, the plasma beta inside the jet-like outflow is generally low  except for the region close to the rotational axis. It implies that the jet-like outflow is generated and accelerated mainly by the magnetic field.  In the counter-rotating case, on the other hand, the plasma beta inside the jet-like outflow is complicated. The plasma beta is low near the central black hole. At the edge of the jet-like outflow, the plasma beta is high. When the black hole rotation is fast, the magnetic field is twisted counter to the rotation of stellar matter in the counter-rotating case. Hence, the direction of the toroidal magnetic field generated near the black hole is opposite to that in the accreting matter far from the black hole. Such counter twisted magnetic field propagates outward to collide and release magnetic twist in the accreting matter in the distant region. As a result of this releasing, the region of weak toroidal magnetic field is produced locally. This region becomes high plasma beta region. This high plasma beta region is also seen in the counter-rotating case of Koide et al. (2000) which simulated the rotating black hole with a Keplerian accretion disk by using a general relativistic MHD code. This high plasma beta region seems inherent to the counter-rotating case.

For the rotating black hole, we cannot see a gas-pressure-driven jet that was found in the general relativistic MHD simulations of a black hole (non-rotating and rotating) with an accretion disk (Koide, Shibata, \& Kudoh 1998, 1999, Koide et al. 2000; Aoki et al. 2004). The gas-pressure-driven jet is generated by the shock produced in the equatorial plane of an accretion disk by the centrifugal barrier. Since we assume in our simulations almost rigid rotation, the centrifugal barrier is not so effective in the initial accretion phase. The matter falls to the central black hole smoothly without the generation of a shock. If our simulations run long enough to form an accretion disk around the central black hole, the gas-pressure-driven jet may be generated.

\subsection{Properties of the Jet}

\begin{figure}
\plotone{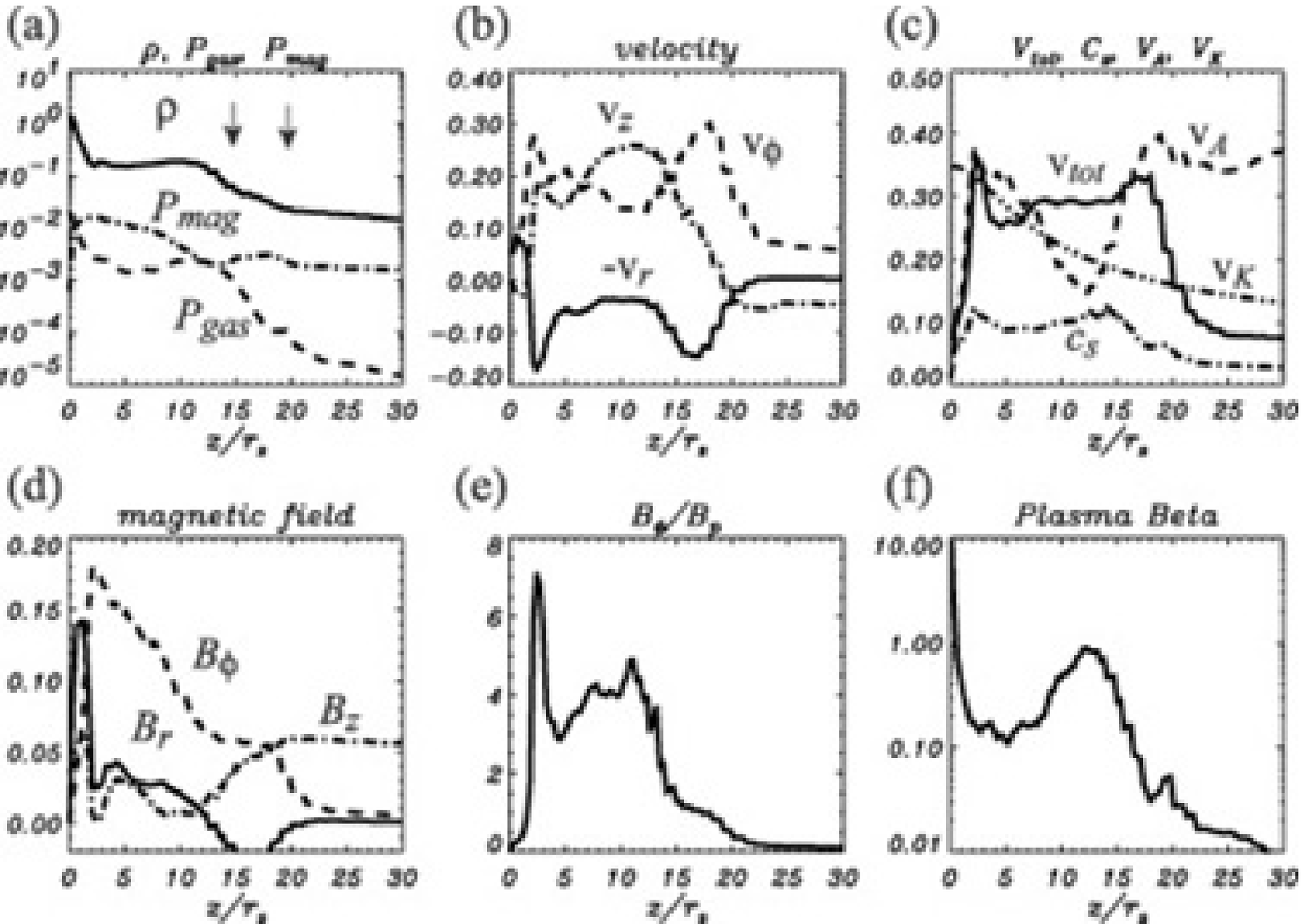}
\caption{The various physical quantities on the surface of $x/r_{\mathrm{S}} = 5$ at $t/\tau_{\mathrm{S}} = 136$ for the co-rotating case KA1. (a) Density $\rho$ (solid line), gas pressure $P_{\mathrm{gas}}$ (dashed line), and magnetic pressure $P_{\mathrm{mag}}$ (dot-dashed line). Allows represent the location of shocks. (b) The components of velocity, $v_{r}$ (solid line), $v_{\phi}$ (dashed line), and $v_{z}$ (dot-dashed line). (c) The total velocity $v_{\mathrm{tot}} = \sqrt{v_{r}^{2}+ v_{\phi}^{2} + v_{z}^{2}}$ (solid line), Alfv\'{e}n velocity $v_{\mathrm{A}}$ (dashed line), sound velocity $c_{\mathrm{s}}$ (dot-dashed line), and Keplerian velocity $v_{\mathrm{K}}$ (double-dot-dashed line). (d) The components of the magnetic field, $B_{r}$ (solid line), $B_{\phi}$ (dashed line), and $B_{z}$ (dot-dashed line). (e) The ratio of toroidal to poloidal magnetic field components $ B_{\phi} / B_{p}$. (f) the plasma beta $\beta = P_{\mathrm{gas}}/ P_{\mathrm{mag}}$. The maximum jet velocity is mildly-relativistic ($\sim 0.3$ c). However, it is larger than the escape velocity. \label{fig3}}
\end{figure}

\begin{figure}
\plotone{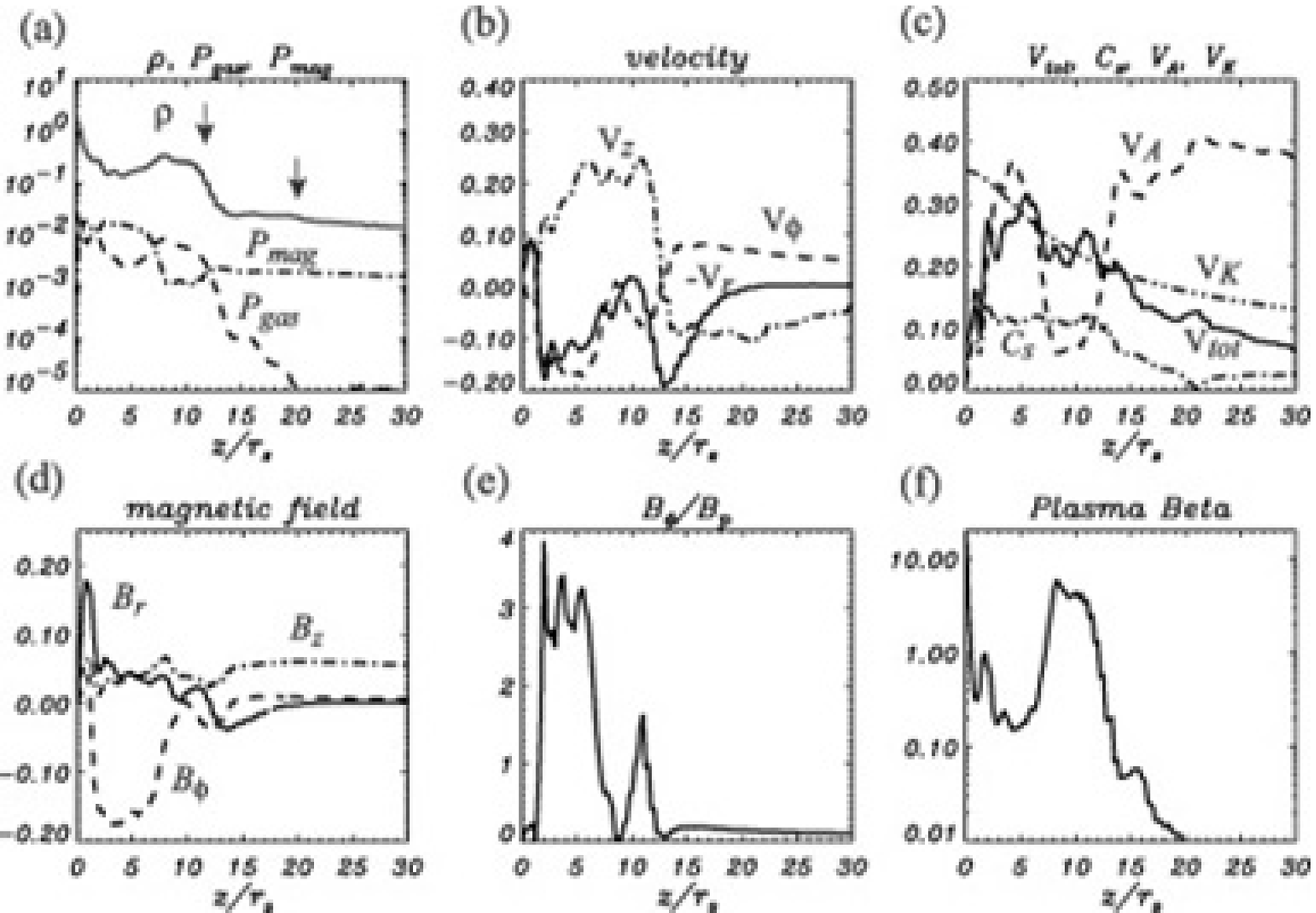}
\caption{The various physical quantities on the surface of $x/r_{\mathrm{S}} = 5$ at $t/\tau_{\mathrm{S}} = 136$ for the counter-rotating case KA2. (a) Density $\rho$ (solid line), gas pressure $P_{\mathrm{gas}}$ (dashed line), and magnetic pressure $P_{\mathrm{mag}}$ (dot-dashed line). Allows represent the location of shocks. (b) The components of velocity, $v_{r}$ (solid line), $v_{\phi}$ (dashed line), and $v_{z}$ (dot-dashed line). (c) The total velocity $v_{\mathrm{tot}}$ (solid line), Alfv\'{e}n velocity $v_{\mathrm{A}}$ (dashed line), sound velocity $c_{\mathrm{s}}$ (dot-dashed line), and Keplerian velocity $v_{\mathrm{K}}$ (double-dot-dashed line). (d) The components of the magnetic field, $B_{r}$ (solid line), $B_{\phi}$ (dashed line) and $B_{z}$ (dot-dashed line). (e) The ratio of toroidal to poloidal magnetic field components $ B_{\phi} / B_{p}$. (f) The plasma beta $\beta = P_{\mathrm{gas}}/ P_{\mathrm{mag}}$. The maximum jet velocity is mildly-relativistic ($\sim 0.25$ c). It is comparable to the escape velocity.\label{fig4}}
\end{figure}

We discuss here the properties of the jet found in the rotating black hole case. Figures \ref{fig3} and \ref{fig4} show the distributions of various physical quantities along the jet-like outflow at $t/\tau_{\mathrm{S}}=136$ in the co-rotating case KA1 and in the counter-rotating case KA2, respectively. 

It is easily seen in the velocity distribution that the jet has a mildly relativistic velocity, $\sim 0.3c$ in the co-rotating case and $\sim 0.25c$ in the counter-rotating case. It is supersonic in both cases. The total velocity of jet in the co-rotating case is clearly larger than the Keplerian velocity. 
This means that the jet-like outflow in the co-rotating case is likely to get out of the stellar remnant. On the other hand, the total velocity of jet for the counter-rotating case is comparable to the Keplerian velocity, so that the jet in this case may not get out of the stellar remnant. Although the total velocity of the jet for the rotating black hole is comparable to that for the non-rotating black hole, the poloidal velocity of the jet in the former case is about 3 times higher than in the latter. The poloidal velocity is the dominant component of the velocity in the jet-forming region for both co- and counter-rotating cases. This is contrary to the result for the non-rotating black hole. Because the jet-like outflow is produced from the deeper gravitational potential region for the rotating black hole cases than for the non-rotating black hole, more gravitational energy is released as a kinetic energy of the jet. 

It is seen that two shock waves propagate outward in both cases. One is a fast shock located in about 20 $r_{\mathrm{S}}$. The other is a slow shock located in about 15 $r_{\mathrm{S}}$ (see Fig. {\ref{fig11}}). Across the fast shock magnetic field strength increase, while across the slow shock magnetic field strength decrease. 
The density of the jet is about an order of magnitude higher than that of the surrounding matter. The toroidal magnetic field ($B_{\phi}$) is the dominant component of the magnetic field in the jet-forming region. This is the same as for the non-rotating black hole case. The ratio of $B_{\phi}$ to $B_{p}$ for the rotating black hole is about 3 times as high as that for the non-rotating black hole. When the magnetic twist is large, magnetic pressure becomes large. The large magnetic pressure pushes up the accreting stellar matter as the jet-like outflow rather than rotates along the magnetic field. Therefore the vertical component of velocity becomes dominating. 

\begin{figure}
\plotone{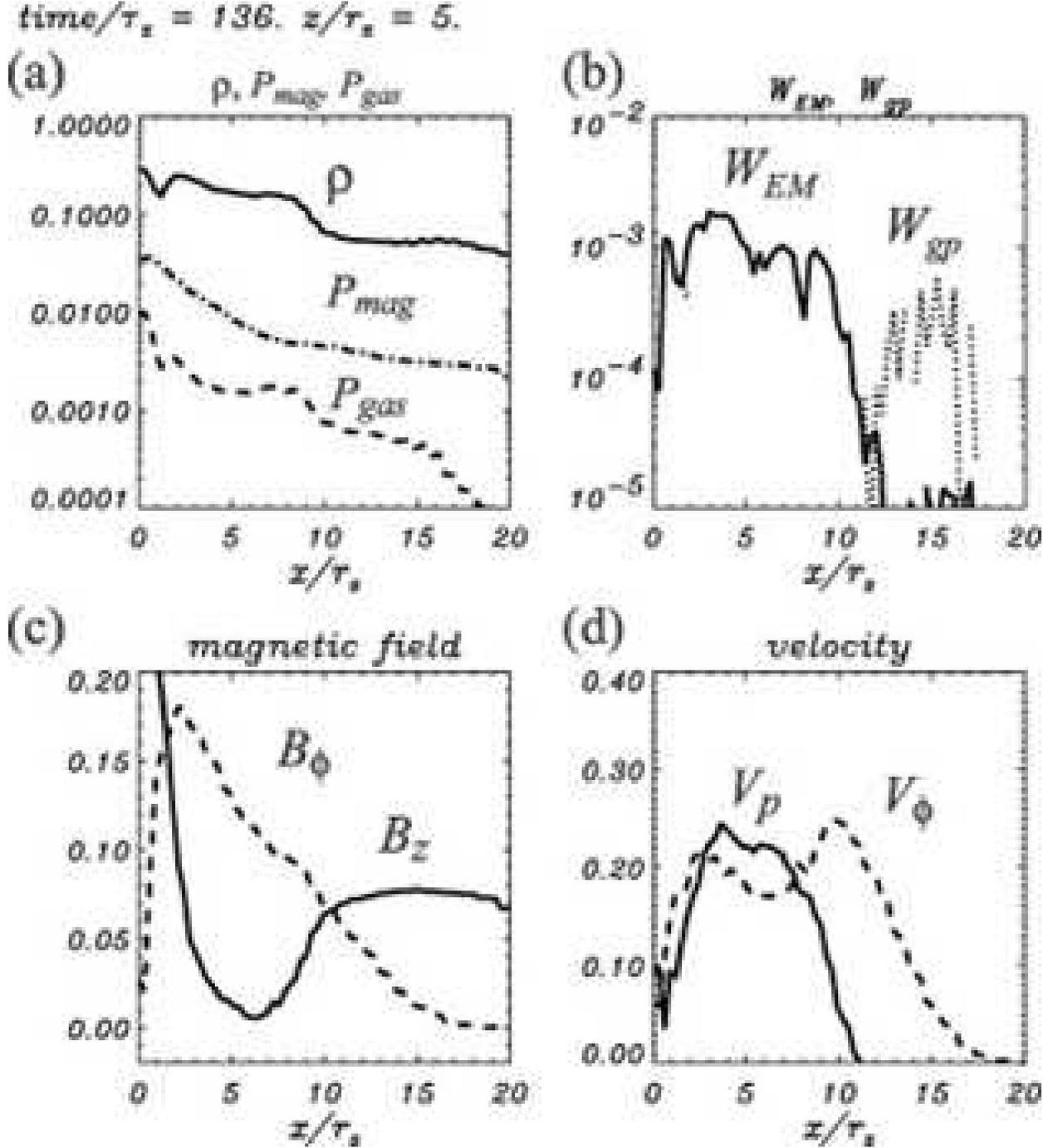}
\caption{The various physical quantities on the surface of $z/r_{\mathrm{S}} = 5$ at $t/\tau_{\mathrm{S}} = 136$ for the co-rotating case KA1. (a) Density $\rho$ (solid line), gas pressure $P_{\mathrm{gas}}$ (dashed line), and magnetic pressure $P_{\mathrm{mag}}$ (dot-dashed line). (b) The power contribution of the vertical component of the electromagnetic force $W_{\mathrm{EM}}$ (solid line) and the gas pressure $W_{\mathrm{gp}}$ (dashed line). (c) The components of the magnetic field, $B_{z}$ (solid line) and $B_{\phi}$ (dashed line). (d) The components of velocity, $v_{p}$ (solid line) and $v_{\phi}$ (dashed line). The jet is located in the region $1.0 r_{\mathrm{S}} < x < 10 r_{\mathrm{S}}$. The edge of the expanding amplified magnetic field is located in the region $10 r_{\mathrm{S}} < x < 15 r_{\mathrm{S}}$. \label{fig5}}
\end{figure}

We show the distributions of various physical quantities on the surface of $z/r_{\mathrm{S}} = 10$ at $t/\tau_{\mathrm{S}}=136$ for the co-rotating case KA1 (Fig.\ref{fig5}). In order to confirm the jet acceleration mechanism, we evaluate the power, $W_{\mathrm{EM}}$, by the electromagnetic force and that, $W_{\mathrm{gp}}$, by the gas pressure as
\begin{equation}
W_{EM} \equiv \mathbf{v} \cdot (\mathbf{E} + \mathbf{J} \times \mathbf{B}),
\end{equation}
\begin{equation}
W_{gp} \equiv - \mathbf{v} \cdot \nabla p,
\end{equation}
respectively. The jet-like outflow in the co-rotating case is located in the region $1.0 r_{\mathrm{S}} < r < 10 r_{\mathrm{S}}$. In this region, the density and the pressure are higher than those of the surrounding region, and the toroidal magnetic field and the vertical velocity are the dominant components of magnetic field and velocity, respectively. The jet-like outflow in the co-rotating case is mainly accelerated by the electromagnetic force because the electromagnetic force is high in this region. Hence, it is a magnetically driven jet. The edge of the expanding amplified magnetic field is located in the region $10 r_{\mathrm{S}} < r < 15 r_{\mathrm{S}}$ outside the jet. In this region, the toroidal velocity is the dominant component of the velocity and the power of gas pressure is high. 

\begin{figure}
\plotone{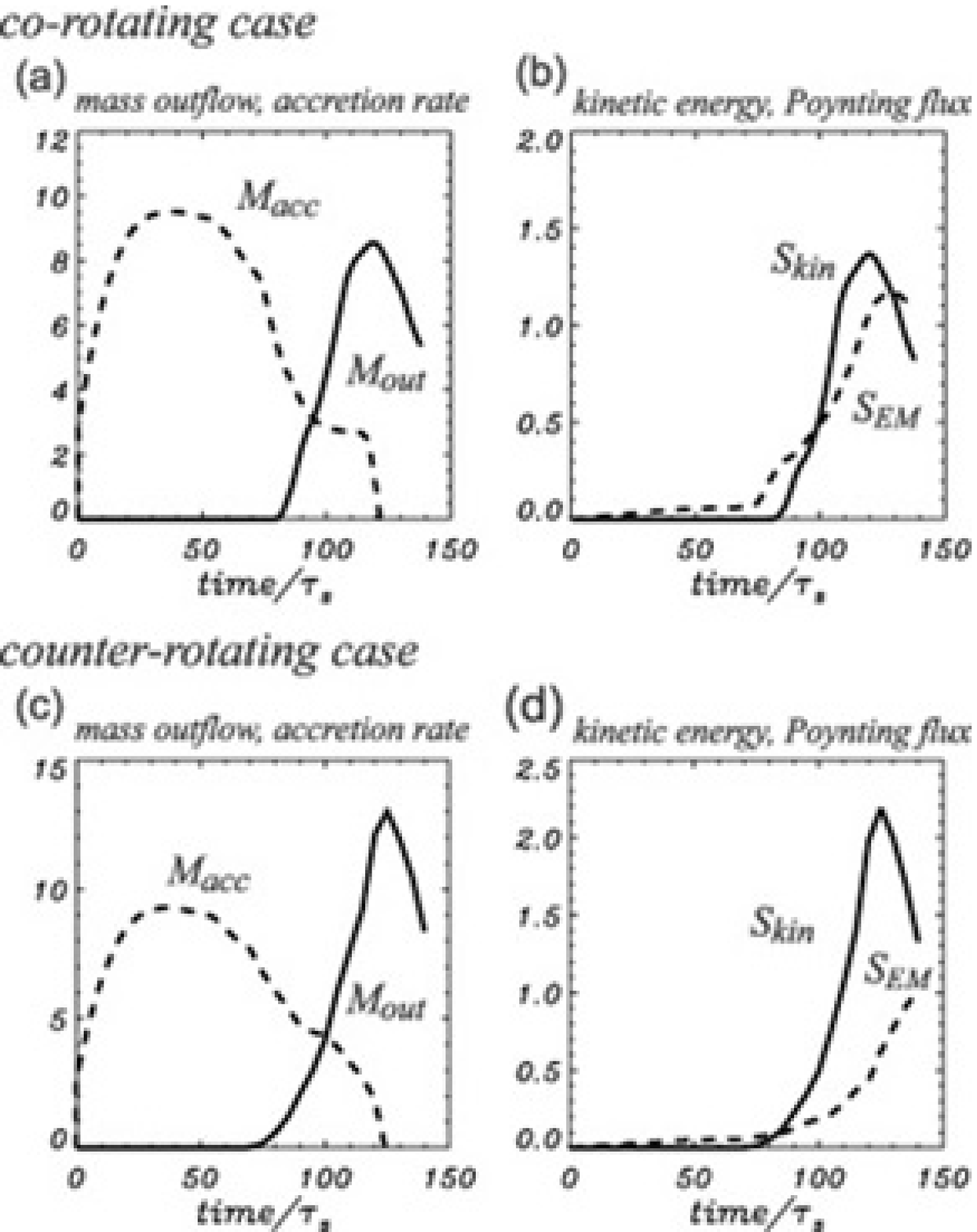}
\caption{The time variation of the mass flux of (a), (c) the jet (solid line) and the accretion rate (dashed line), (b), (d) the kinetic energy (solid line) and the Poynting flux (dashed line) at $z/r_{\mathrm{S}} \simeq 12$ for the co-rotating case KA1 (upper panel) and the counter-rotating case KA2 (lower panel). \label{fig6}}
\end{figure}

We show the time variations of the mass flux of the jet, accretion rate, kinetic energy and Poynting flux at $z/r_{\mathrm{S}} \simeq 15$ for the co-rotating case KA1 and the counter-rotating case KA2 in Figure \ref{fig6}. In the co-rotating case, the kinetic energy flux is comparable to the Poynting flux, while, in the counter-rotating case, the kinetic flux is about twice as large as the Poynting flux. These results differ from those for the non-rotating black hole. This is because the jet-like outflow for the rotating black hole is faster in the vertical direction and denser than for the non-rotating black hole. The Poynting flux for the rotating black hole is also larger, because the magnetic field is twisted strongly due to the frame dragging effect of the rotating black hole.

\subsection{Dependence on the rotation parameter}

\begin{figure}
\plotone{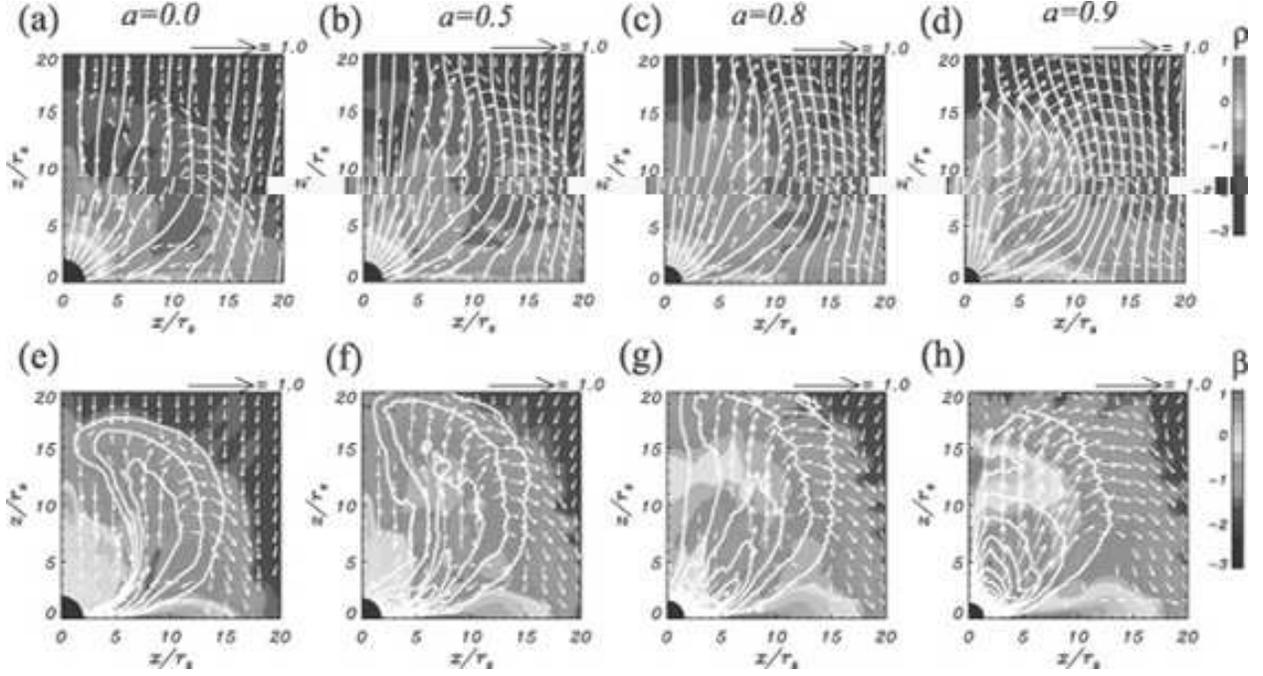}
\caption{The snapshots of density and plasma beta in the co-rotating cases having different rotation parameter KB3 ($a=0.0$)(a, e), KB5 ($a=0.5$)(b, f), KB6 ($a=0.8$)(c, g), and KB7 ($a=0.9$)(d, h) at $t/ \tau_{\mathrm{S}} = 136$. The color scales show the values of the logarithm of density and plasma beta. The white curves represent the magnetic field lines (a, b, c, d) and the contour of the toroidal magnetic field (e, f, g, h). The contour level step-width is 0.025 for (e), (f), and (g), 0.05 for (h) in units of the toroidal magnetic field ($B_{\phi}$). Arrows depict the poloidal velocities normalized by the light velocity. For smaller values of the rotation parameter, the jet is ejected more from the outer region and the expansion of the amplified magnetic field as an Alfv\'{e}n wave is seen more clearly. \label{fig7}}
\end{figure}

The dependence of the jet properties on the rotation parameter of the black hole has been investigated. Figure \ref{fig7} shows the snapshots of density and plasma beta in the co-rotating cases with different rotation parameters, $a=0.0$ (KB3), $a=0.5$ (KB4), $a=0.8$ (KB5) and $a=0.9$ (KB6) at $t/ \tau_{\mathrm{S}} =136$. The difference between KB4 ($a=0.5$) and KB6 ($a=0.8$) is not seen clearly. On the other hand, the difference between KB6 ($a =0.8$) and KB7 ($a = 0.9$) is clear. For smaller values of the rotation parameter, the jet is ejected from more outer regions and the propagation of the amplified magnetic field as Alfv\'{e}n waves is faster and is seen more clearly. This implies that the inner magnetic field is amplified strongly by the frame dragging effect.

\begin{figure}
\plotone{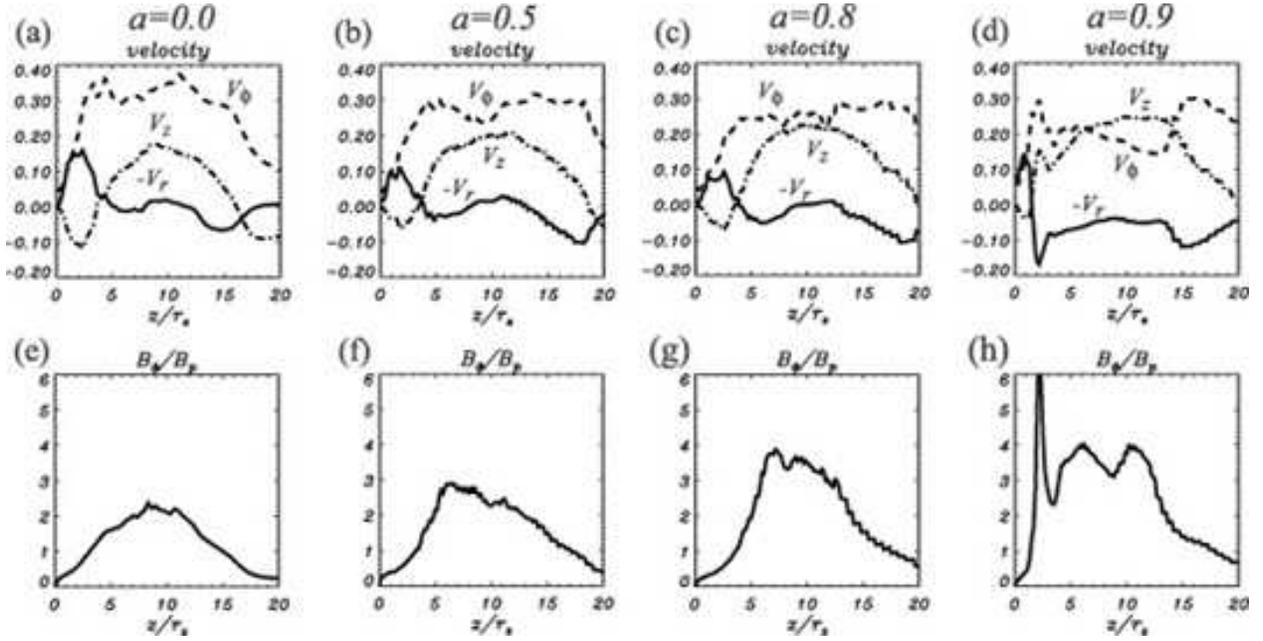}
\caption{The distribution of velocity ($v_{r}$, $v_{\phi}$ and $v_{z}$) and the ratio of toroidal to poloidal magnetic field components ($B_{\phi}/B_{p}$) along the jet in the co-rotating cases having different rotation parameter KB3($a=0.0$)(a,e), KB5 ($a=0.5$)(b, f), KB6 ($a=0.8$)(c, g), and KB7 ($a=0.9$)(d, h) at $t/ \tau_{\mathrm{S}} = 136$. For smaller values of the rotation paramter, the toroidal component of velocity is larger, the poloidal component of velocity is smaller and the ratio of toroidal to poloidal magnetic field components becomes smaller. \label{fig8}}
\end{figure}

Figure \ref{fig8} shows the distribution of velocity ($v_{r}$, $v_{\phi}$ and $v_{z}$) and the ratio of toroidal to poloidal magnetic field components ($B_{\phi}/B_{p}$) along the jet in the co-rotating cases with different rotation parameters $a=0.0$ (KB3), $a=0.5$ (KB4), $a=0.8$ (KB5) and $a=0.9$ (KB6) at $t/ \tau_{\mathrm{S}} = 136$. The differences between cases are apparent. As the rotation of the black hole becomes slower, the toroidal component of the velocity becomes faster and the poloidal component of the velocity becomes slower. The ratio of toroidal to poloidal magnetic field components gets larger as the rotation of black hole becomes slower. As the rotation of the black hole is faster, the magnetic field is twisted more strongly in shorter times owing to the frame-dragging effect. Hence, a stronger and faster jet is produced near the central black hole. 

\begin{figure}
\epsscale{0.35}
\plotone{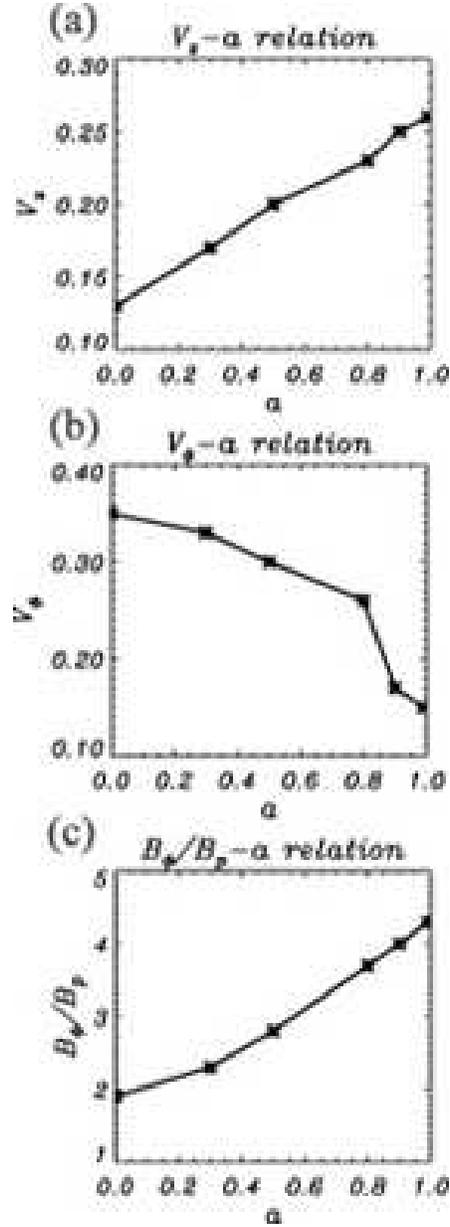}
\caption{Dependence of (a) maximum vertical velocity of the jet ($v_{z}$), (b) the maximum toroidal velocity of the jet ($v_{\phi}$) and (c) the maximum ratio of toroidal to poloidal magnetic field components ($B_{\phi}/B_{p}$) of the jet on the rotation parameter of a black hole, $a$, in the co-rotating case. \label{fig9}}
\end{figure}

The dependence on the rotation parameter of the black hole in the co-rotating case is shown in Figure \ref{fig9}. As the rotation parameter of the black hole increases, the poloidal velocity of the jet and the magnetic twist increase gradually and the toroidal velocity of the jet decreases. These results are understood from how effective the frame-dragging effect is. As the rotation of the black hole is faster, the magnetic energy stored by the twisted magnetic field is converted to kinetic energy of the jet more directly rather than to the propagation of Alfv\'{e}n waves. As a result, the poloidal velocity of the jet becomes higher.
Let us examine the above statement more quantitatively. In the Newtonian approximation, the time evolution of the toroidal magnetic field is given as
\begin{equation}
{\partial B_{\phi} \over \partial t} \sim \omega B_{p},
\end{equation}
where $\omega$ is the angular velocity, which includes the rotation of both matter and the frame (space-time),
\begin{equation}
\omega \propto a.
\end{equation}
From this, we obtain
\begin{equation}
{B_{\phi} \over B_{p}} \sim \omega t \propto a.
\end{equation}

The upward motion of the fluid is induced by the $\mathbf{J} \times \mathbf{B}$ force. If we neglect other forces, the equation of motion for the fluid element in the z-direction becomes
\begin{equation}
\rho {\partial v_{z} \over \partial t} \sim \nabla \left( {B_{\phi}^{2} \over 4 \pi} \right) \sim {1 \over z} \left( {B_{\phi}^{2} \over 4 \pi} \right),
\end{equation}
which can be rewritten as
\begin{equation}
v_{z} \sim {1 \over \rho} {t \over z} \left( {B_{\phi}^{2} \over 4 \pi} \right).\label{eq2-20}
\end{equation}
In the current situation, the time scale is given by the propagation time of the Alfv\'{e}n wave and the Alfv\'{e}n velocity is written as $v_{A} \sim v_{A \phi} = B_{\phi}/ \sqrt{4 \pi \rho}$ because $B_{\phi}/B_{p} > 1$. Thus, we have $z/t \sim v_{A \phi}$. Then, equation (\ref{eq2-20}) can be rewritten as
\begin{equation}
v_{z} \sim {1 \over \rho}{1 \over v_{A \phi}} \left( {B_{\phi}^{2} \over 4 \pi} \right) \propto B_{\phi} \propto a.
\end{equation}
This explains the dependence of $v_{z}$ on $a$ in Figure \ref{fig9}(a), that is,  the vertical component of the jet velocity increases as the initial magnetic field strength increases.

On the other hand, the rotation of the fluid is also induced by the $\mathbf{J} \times \mathbf{B}$ force. The equation of motion for the fluid element in the toroidal direction becomes
\begin{equation}
\rho {\partial v_{\phi} \over \partial t} \sim \nabla \left( {B_{\phi} B_{z} \over 4 \pi} \right) \sim {1 \over z} \left( {B_{\phi} B_{z} \over 4 \pi} \right).
\label{eq2-21}
\end{equation}
Using $z/t \sim v_{A \phi}$ in Eq. (\ref{eq2-21}) leads to
\begin{equation}
v_{\phi} \sim {1 \over \rho}{1 \over v_{A \phi}} \left( {B_{\phi} B_{z} \over 4 \pi} \right) \propto B_{z} = const.
\end{equation}
This approximately explains the dependence of $v_{\phi}$ on $a$  for $a < 0.8$ in Figure \ref{fig9}(b). However, the exact relation depends on the region where the jet is ejected. In fact, the jet ejected from a deeper region has a lower toroidal velocity owing to the conservation of angular momentum.

\begin{figure}
\epsscale{1.0}
\plotone{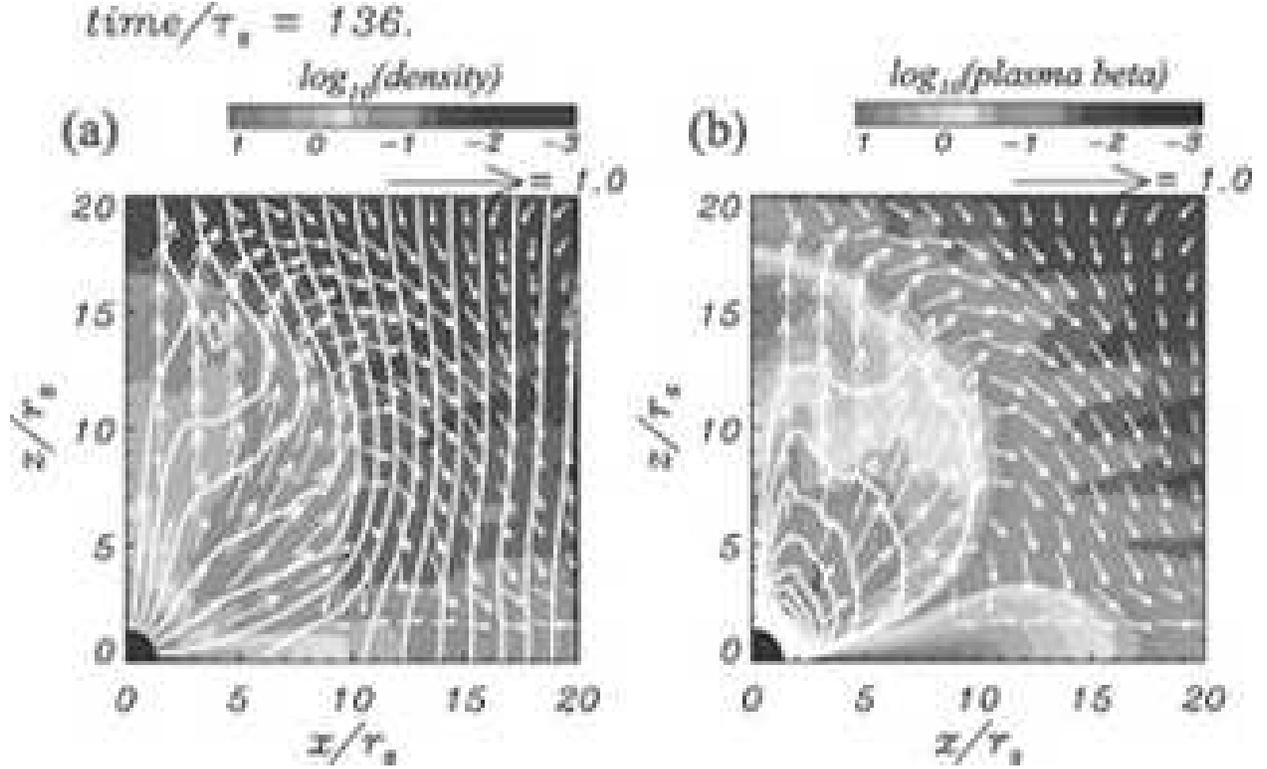}
\caption{The snapshots of density (a) and plasma beta (b) for the {\it no stellar rotation} case KC8 at $t/\tau_{\mathrm{S}} = 136$. The color scales show the values of the logarithm of density and plasma beta. The white curves represent the magnetic field lines (a) and the contour of the toroidal magnetic field (b). The contour level step-width is 0.067 for (b) in units of the toroidal magnetic field ($B_{\phi}$). Arrows depict the poloidal velocities normalized by the light velocity. Although there is no stellar rotation, the jet-like flow is produced near the black hole. \label{fig10}}
\end{figure}

We computed the case KC8 in which {\it the accreting matter is not rotating} to investigate the effect of frame dragging. Figure \ref{fig10} shows the snapshots of density and plasma beta at $t/\tau_{\mathrm{S}} = 136$. The simulation results show that the jet-like outflow can be formed solely by the frame dragging effect. 
Thus the rapid rotation of black hole will help to form the jet-like outflow due to the twisting of magnetic field by the frame dragging effect. We also see a disk-like structure on the equatorial plane, which is {\it not} rotating. Hence this disk is not an accretion disk. The disk-like structure is produced by the effect of magnetic field and supported by the gas pressure. All disk-like structures seen in other cases are basically the same as this non-rotating disk, and thus are not an accretion disk.

\section{Summary and Discussion}

\begin{figure}
\plotone{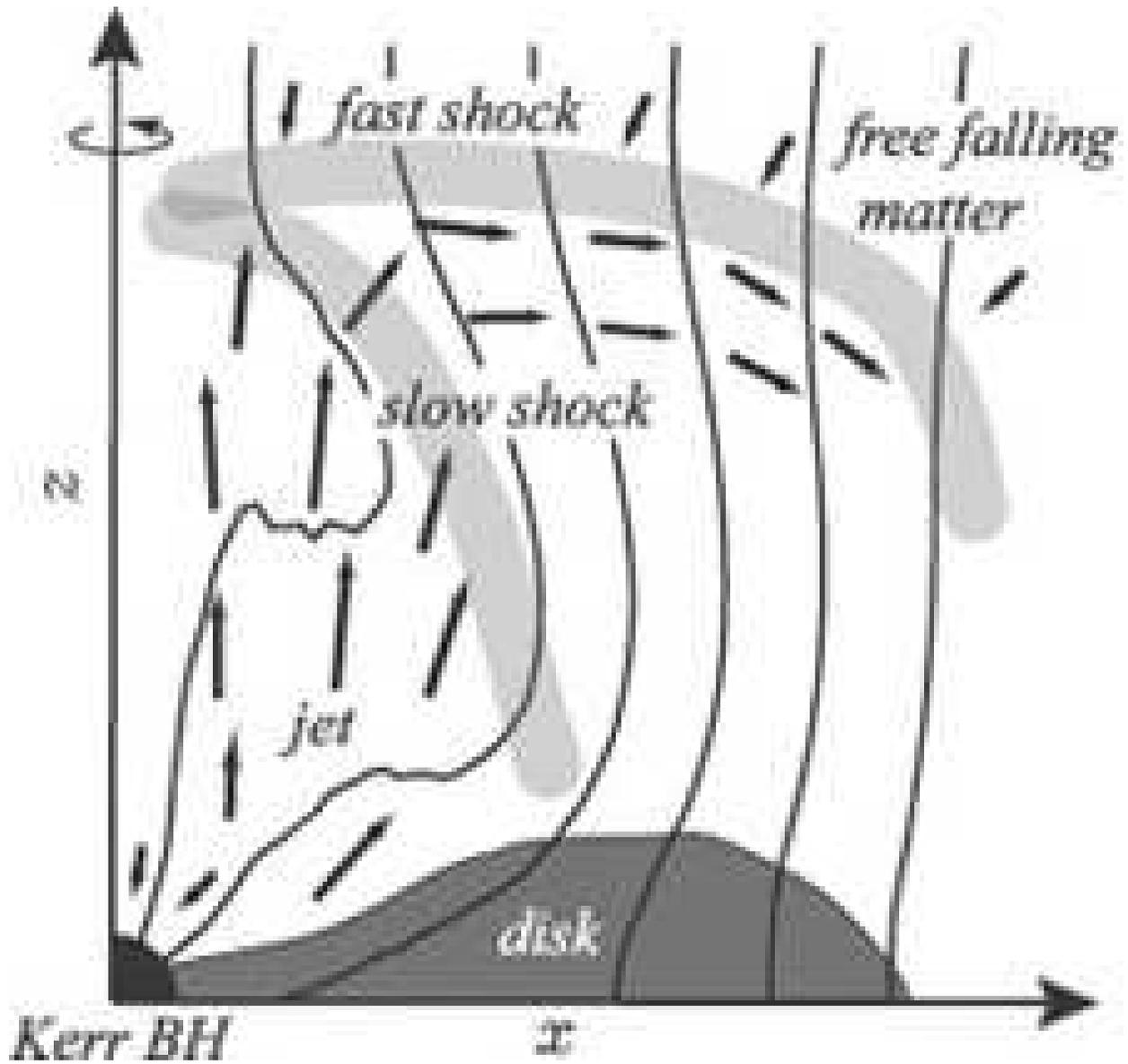}
\caption{Schematic picture of our sumulation results. A disk-like structure and a jet-like outflow are formed in the vicinity of the black hole. \label{fig11}}
\end{figure}

We have performed general relativistic MHD simulations of collapsar, paying particular attention to the rotation of the central black hole. We considered not only the co-rotating but also the counter-rotating cases with respect to the black hole rotation to elucidate its effect on the dynamics. Our results are summarized as follows: 

\begin{enumerate}
	\item The formation mechanism of the jet for the rotating black hole is the same as that for the non-rotating black hole. When stellar matter falls onto the black hole, a disk-like structure is formed in the vicinity of the black hole horizon and a jet-like outflow is formed near the black hole mainly by the magnetic field. The jet-like outflow propagates outward with twisted magnetic fields and becomes collimated. The shock waves (MHD fast shock and slow shock) are also formed near the black hole and propagate outward. Figure {\ref{fig11}} shows the schematic picture of our simulation results.

     \item The total velocity of jet for the rotating black hole is comparable to that for the non-rotating black hole, $\sim 0.3c$. However, the poloidal velocity of the jet in the case of rotating black hole is about 3 times higher than that in the case of non-rotating black hole. The poloidal velocity is the dominant component of the jet. Because the magnetic field is more strongly twisted due to the frame dragging effect of the rotating black hole, more magnetic energy is converted to the kinetic energy of the jet rather than to the propagation of Alfv\'{e}n waves.    
 
    \item In the co-rotating case, the kinetic energy flux is comparable to the Poynting flux. In the counter-rotating case, on the other hand, the kinetic energy flux is about twice as large as the Poynting flux. This is because the jet for the rotating black hole has a higher density. 

    \item As the rotation parameter of the black hole increases, the poloidal velocity of the jet and the magnetic twist increase gradually and the toroidal velocity of the jet decreases. These results are related to how effective the frame-dragging is. As the rotation of black hole is faster, the magnetic field is twisted more strongly owing to the frame-dragging effect. The magnetic energy stored by the twisted magnetic field is converted to the kinetic energy of the jet directly rather than leading to the propagation of Alfv\'{e}n waves. Thus, the poloidal velocity of the jet becomes faster as the rotation of the black hole becomes faster.

   \item In rapidly rotating black hole case the jet-like outflow can be formed by the frame dragging effect through the twisting of magnetic field even if there is no stellar rotation.
   
\end{enumerate}

We discuss the application of our results to the central engine of GRBs. 
Using the simulation data, we can estimate the kinetic energy of the jet, $E_{\mathrm{jet}}$. If the mass of the disk formed in the initial accretion phase is comparable to the central black hole mass, e.g., several solar masses (almost all stellar matter except the central core are actually blown away by the presupernova), the density of the disk can be estimated from 
\begin{equation}
 \pi r_{\mathrm{d}}^{2} H \rho_{\mathrm{d}} = 4 M_{\odot} \simeq 10^{34} \mathrm{g} \label{},
\end{equation}
where $r_{\mathrm{d}}$ is the outer radius, $H$ the thickness and $\rho_{\mathrm{d}}$ the density of the disk, respectively. $r_{\mathrm{d}}$ and $H$ can be estimated from the simulation data as $r_{\mathrm{d}} = 10 r_{\mathrm{S}} \simeq 10^{7} \mathrm{cm}$, $H = 0.1 r_{\mathrm{d}} = 1 r_{\mathrm{S}} \simeq 10^{6} \mathrm{cm} $, respectively. Therefore, the density of the disk is $ \rho_{\mathrm{d}} \simeq 3 \times 10^{13} \mathrm{g/cm^{3}} $. Because the jet-like outflow for the rotating black hole is generated very close to the central black hole, the density of the jet is comparable to the density of the disk. Thus, using the estimated density of the disk, the kinetic energy of the jet is given as $E_{\mathrm{jet}} \simeq 10^{54}$ ergs. Although it may somewhat overestimate, this is large enough energy to explain the standard energy ($\sim 10^{51}$ ergs) of GRBs .
However, the maximum jet velocity in our simulations including a rotating black hole was about $0.3 c$. This velocity is too slow compared to the velocity inferred for the jet of GRBs. We have to consider other acceleration mechanisms. This is the most difficult and fundamental problem in all types of ultrarelativistic outflow, e.g. AGN jets, pulsar winds and GRBs. 
In order to obtain not only a large energy extraction but also the observed large bulk Lorentz factors, the Poynting flux must be converted to the kinetic energy. From steady solution of relativistic outflow in ideal MHD, it is found that  the energy conversion from the Poynting flux to the kinetic energy flux occurs and the outflow is highly accelerated, if the magnetic field lines diverge with radius more quickly than in the monopole field (Begelman \& Li 1994; Takahashi \& Shibata 1998; Daigne \& Drenkhahn 2002). However, this solution is not self-consistent because the geometry of the magnetic field is not solved. Moreover, the relativistic outflow may not maintain the collimated structure in this situation. Some authors proposed a dissipation-induced acceleration mechanism (Spruit, Daigne, \& Drenkhahn 2001; Drenkhahn 2002; Drenkhahn \& Spruit 2002; Sikora et al. 2003). Although the centrifugal acceleration may also contribute to the early stages of acceleration of the flow, this is not a magneto-centrifugal acceleration process. If the magnetic fields in the outflow change the direction on sufficiently small scales, a part of the magnetic energy can be released locally by magnetic reconnections. This process has two useful effects. First, it converts Poynting flux directly into radiation, without only intermediate step with internal shocks. Second, it leads to steeper decline of magnetic pressure, which causes great acceleration of the flow and enhanced conversion of the Poynting flux to the kinetic energy. Since in the co-rotating case of our simulations, the Poynting flux is comparable to the kinetic energy flux, the dissipation-induced acceleration mechanism which might occur later on may be the key to solve the acceleration problem of our simulations. 

Although it may not be a main solution of the acceleration problem, we further discuss possible acceleration mechanisms from other simulations of relativistic jets. The first possibility is a disk-jet, especially a gas-pressure-driven jet. The gas-pressure-driven jet was seen in general relativistic MHD simulations of the black hole-accretion disk system (Koide et al. 1998, 1999, 2000; Aoki et al. 2004) and has higher velocity than the magnetically driven jet. It is generated by the shock produced by the centrifugal barrier in the equatorial plane. Since, in our simulations, we initially assumed that the rotation is almost rigid, the centrifugal barrier is not so effective in the initial accretion phase. The accreted matter falls to the central black hole smoothly without the generation of a shock. We expect that if our simulations run long enough to form the accretion disk around the central black hole, the gas-pressure-driven jet will be generated. The second possible mechanism is the break out of the jet from the stellar surface. When the jet emerges from the stellar surface, the steep density gradient accelerates the jet. Some authors (Aloy et al. 2000; Zhang et al. 2003) have showed numerically that the jet is accelerated significantly by this process, and the terminal Lorentz factor becomes more than 100. Since the jet in our simulations contains large energy, this mechanism may work effectively. We will address the propagation of the jet outside the stellar surface in a forthcoming paper.

Our results can be also applied to baryon-rich outflows associated with failed GRBs. The baryon-rich outflow is a fireball with high baryonic load and mildly relativistic velocities. The jet velocity is so slow that it cannot produce GRBs. Such failed GRBs are supposed to occur at higher event rates than successful GRBs (Woosley et al. 2002; Huang, Dai, \& Lu 2002).  Some failed GRBs may be observed as ``hypernovae''. Particularly, SN 2002ap is a candidate for such an event. Although the association with a GRB has not been found , it is inferred to have a jet with a velocity of $\sim 0.23 c$ and estimated kinetic energy of $\sim 5 \times 10^{50} \mathrm{ergs}$ (Kawabata et al. 2002; Totani 2003). The jet found in our simulations for rotating black holes has comparable velocity and large enough energy to explain the jet of SN2002ap.  

In our simulations, we have neglected microphysics (photodisintegration, treatment of neutrinos etc.) and concentrated on the general relativistic magnetohydrodynamics, particularly the effect of rotating black hole.  Such treatment may be reasonable as a first step. However, microphysics are critically important for realistic GRMHD simulations of collapsar.  We will address these issues in a forthcoming paper.

We have assumed a uniform global magnetic field in the simulations. It is inferred, however, that rotating compact stars in general have a dipole-like magnetic field. Hence, it may be more likely that the rotating stars collapse with a dipole-like magnetic field. Although we think that the assumption of a uniform global magnetic field will not be so bad for the discussion of the generic aspect of collapsars dynamics, the dependence on the configuration of magnetic field is also currently being investigated and will be presented elsewhere.

\medskip

Y. M. appreciates many helpful conversations on GRBs and relativistic effects with S. Aoki, T. Totani, R. Yamazaki, M. Takahashi, and K. Nishikawa. He also thanks K. Uehara, H. Kigure, T. Haugb\o elle. and Y. Kato. This work was partially supported by Japan Science and Technology Cooperation (ACT-JST), Grant-in-Aid for the 21st Century COE ``Center for Diversity and Universality in Physics'' and Grants-in-Aid for the scientific research from the Ministry of Education, Science, Sports, Technology, and Culture of Japan through No.14079202, No.14540226 (PI: K. Shibata) and No.14740166. The numerical computations were partly carried out on the VPP5000 at the Astronomical Data Analysis Center of the National Astronomical Observatory, Japan (yym17b), and partly on the Alpha Server ES40 at the Yukawa Institute for Theoretical Physics of Kyoto University, Japan.

\clearpage

\begin{deluxetable}{lccc}
\tablecolumns{6}
\tablewidth{0pc}
\tablecaption{Models and Parameters \label{table1}}
\tablehead{
\colhead{Case} & \colhead{$a$} & \colhead{$B_{0}$} & \colhead{$v_{0}$}} 
\startdata
KA1 ......&  0.999 & 0.05      &  0.01 \\
KA2 ......& -0.999 & 0.05      &  0.01 \\
KB3 ......&  0.0   & 0.05      &  0.01 \\
KB4 ......&  0.3   & 0.05      &  0.01 \\
KB5 ......&  0.5   & 0.05      &  0.01 \\
KB6 ......&  0.8   & 0.05      &  0.01 \\
KB7 ......&  0.9   & 0.05      &  0.01 \\
KC8 ......&  0.999 & 0.05      &  0.0  \\
\enddata
\tablecomments{Co-rotating case KA1 ($a=0.999$, $B_{0} = 0.05$, $v_{0}=0.01$) and Counter-rotating case KA2 ($a=-0.999$, $B_{0} = 0.05$, $v_{0}=0.01$) are considered to be the standard cases in our simulations. Cases \lq\lq B" differ from case KA1 only in the value of the rotation parameter of the black hole. In Case \lq\lq KC8" there is no stellar rotation.}
\end{deluxetable}

\end{document}